\documentclass[12pt,preprint]{aastex}

\slugcomment{Submitted, ApJ, 17 June 2010}
\shorttitle{New HAT-P-9b Ephemeris}
\shortauthors{Dittmann et al.}

\begin{document}

\title{A Revised Orbital Ephemeris for HAT-P-9b} 

\author{Jason A. Dittmann, Laird M. Close, Louis J. Scuderi, Jake Turner, Peter C. Stephenson}
\affil{Steward Observatory, University of Arizona, Tucson, AZ 85721}

\begin{abstract}
We present here three transit observations of HAT-P-9b taken on 14 February 2010, 18 February 2010, and 05 April 2010 UT from the University of Arizona's 1.55 meter Kuiper telescope on Mt. Bigelow. Our transit light curves were obtained in the I filter for all our observations, and underwent the same reduction process. All three of our transits deviated significantly ($\approx 24$ minutes earlier) from the ephemeris of Shporer et al. (2008). However, due to the large time span between our observed transits and those of Shporer et al. (2008), a $6.5$ second ($2\sigma$) shift downwards in orbital period from the value of Shporer et al. (2008) is sufficient to explain all available transit data. We find a new period of $3.922814 \pm 0.000002$ days for HAT-P-9b with no evidence for significant nonlinearities in the transit period.
\end{abstract}


\section{Introduction}
To date, several hundred extrasolar planets have been discovered around nearby stars. However, transiting extrasolar planets are relatively rare among all known planet discoveries (70 transiting planets out of 431 known planetary systems)\footnote{http://exoplanet.eu}. Planetary transits are unique in that they allow for the direct measurement of the radius of the transiting planet relative to the host star. When combined with radial velocity data and determinations of the stellar properties of the host star, it is possible to determine many bulk properties of the planet. Models of the bulk properties of these planets can be formulated and then compared to observation (for example, Baraffe et al. 2008; Fortney et al. 2007; Burrows et al 2007).

The transit method can also be used to detect additional planets in systems with one transiting object. Deviations from a linear ephemeris by a known transiting extrasolar planet is an indicator of an additional perturbing object in the system. Haghighipour et al. (2008) investigated the effect of an Earth mass companion in various orbital resonances with a hypothetical hot Jupiter in a 3 day orbit around a solar type star. They found that an Earth mass body in a resonant orbit with a hot Jupiter can induce transit timing variations (TTVs) on the order of several minutes in 1:2 interior resonant orbits. Furthermore, Haghighipour et al. (2008) found that these TTVs would be sinusoidal in time as the line of nodes of the planets' orbits precesses.  


Besides TTVs, another indicator of potential perturbing bodies in a planetary system is the maintaining of a nonzero orbital eccentricity over long periods of time despite the tendency of tidal forces from the star attempting to circularize the orbit. While most transiting planets are in nearly circular orbits, there are a small number of planets that do maintain eccentric orbits despite being older than the circularization time scale for the system. The most notable of these systems are the transiting Neptunes GJ 436b (Butler et al. 2004)  and HAT-P-11b (Bakos et al. 2010). GJ 436b has a significantly nonzero eccentricity of $0.15 \pm 0.012$ (Deming et al. 2007), despite being older than its circularization time scale (Maness et al. 2007), which has led to speculation about possible additional (possibly resonant) planets in the system gravitationally ``pumping" the eccentricity. Gravitational interaction with a third body would also lead to transit timing variations as the line of nodes precesses, but searches for this effect have found no evidence for transit timing variations in this system (Pont et al. 2009; Ballard et al. 2010). A search in the similar HAT-P-11 exo-Neptune system has yielded no positive indication for large transit timing variations (Dittmann et al. 2009a). 

HAT-P-9b is a recently discovered $0.78 \pm 0.09 M_{Jup}$, $1.40 \pm 0.06 R_{Jup}$ hot Jupiter in an assumed zero eccentricity, $3.92289 \pm 0.00004$ day orbit about a 12.3 V magnitude F star (Shporer et al. 2008). Due to the relatively large size of the host star ($1.32 \pm 0.07 R_{\sun}$), the transit depth is relatively small, at $\approx 11$ mmag (Shporer et al. 2008). Under the $T_{eq}$ and Safronov number classification investigated by Hansen \& Barman (2007), HAT-P-9b is a Class II planet. Follow up observations were obtained by Shporer et al. (2008) on the 1.2 meter FLWO telescope with the KeplerCam detector with a z filter. The orbital period was determined by the central transit times of four follow up light curves taken over 15 orbital periods at this telescope (Shporer et al. 2008). However, Shporer et al. (2008) note that when they compare the ephemeris of their follow-up light curves with that of their original discovery curve with the HATNet telescopes, there is a minor discrepancy of $1.5 \sigma$ in the transit time, and that further observations should be able to more thoroughly study this possible discrepancy. It was this discrepancy in the period that was the primary motivator for this study. 

\section{Observations \& Reductions}

We observed the transit of HAT-P-9b on 14 February, 2010 UT using the University of Arizona's 61 inch (1.55 meter) Kuiper telescope on Mt. Bigelow, Arizona with the Mont4k CCD, binned 3x3 to 0.43$^{\prime\prime}$/pixel. Our observations were taken with an Arizona-I filter. This filter, when convolved with the response curve of the Mont4k CCD, yields a transmission range very similar to the Cousins-I band.

HAT-P-9 is a F dwarf located at $480 \pm 60$ pc with a V magnitude of 12.3 (Shporer et al. 2008). We utilized exposure times between 16 and 25 seconds, depending on cloud conditions, in order obtain a good balance between image signal to noise and cadence. Read out times on the Mont4k CCD are approximately 10 seconds. Seeing ranged from 1$^{\prime\prime}$.0 to 1$^{\prime\prime}$.3 for all observation nights, which is typical for this site. The conditions during the 14 February observation were photometric, with no moon adding to the sky background. On 18 February, conditions were partially cloudy, and we were only able to obtain a partial transit lightcurve. The light curve consisted of a portion of the middle of transit, and all of egress. Conditions on 05 April were photometric. For each observation, we utilized the same reference stars, placed in the same pixels on the Mont4k CCD, and used the same filter in order to minimize any possible systematic effects. Due to excellent autoguiding, there was less than 5 pixels (2.15\arcsec) of image wander over the course of each data set. The onboard clock is synched to GPS every few seconds in order to ensure accurate absolute time keeping.

\begin{figure}[htp]
\centering
\includegraphics[scale=0.35]{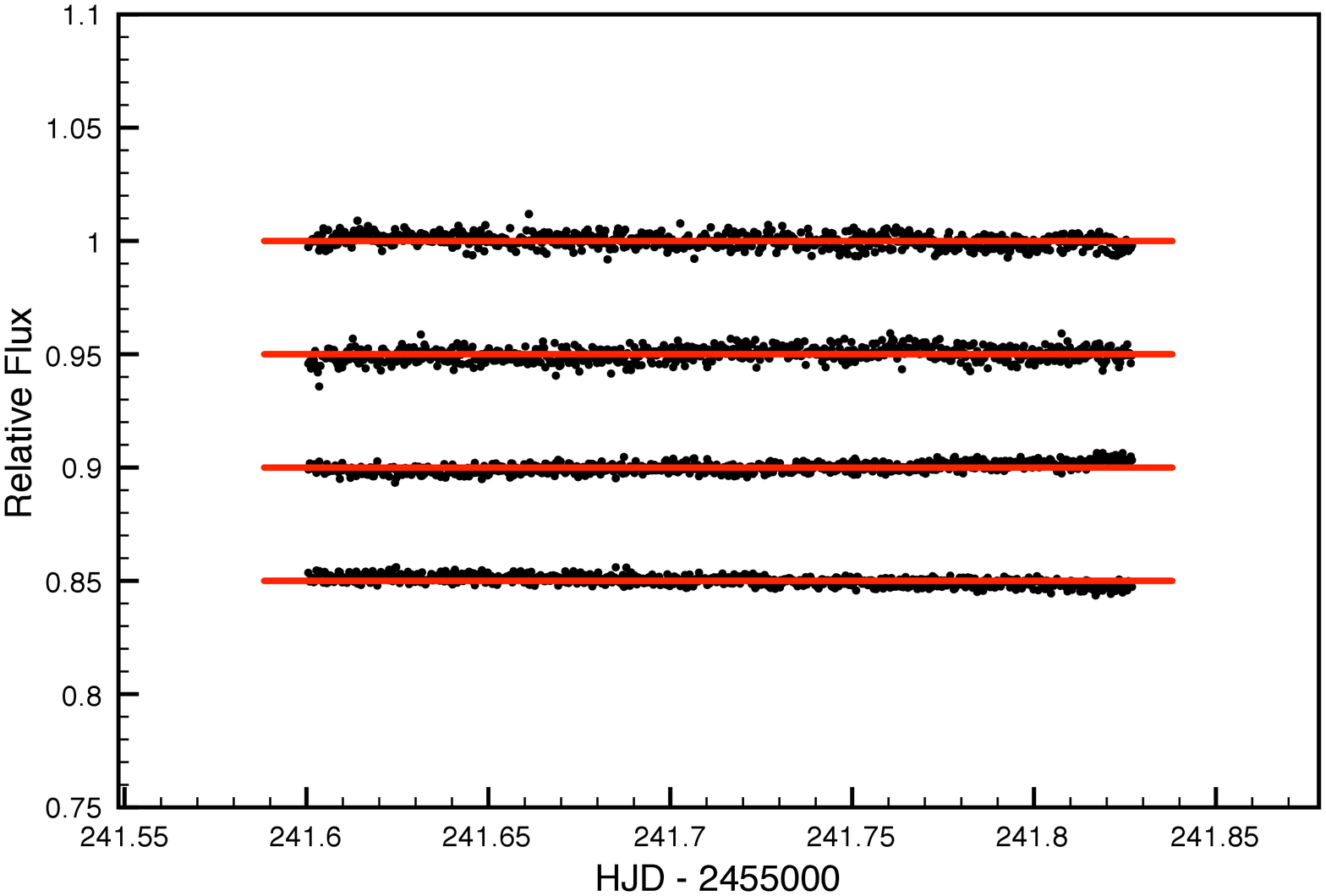}
\includegraphics[scale=0.35]{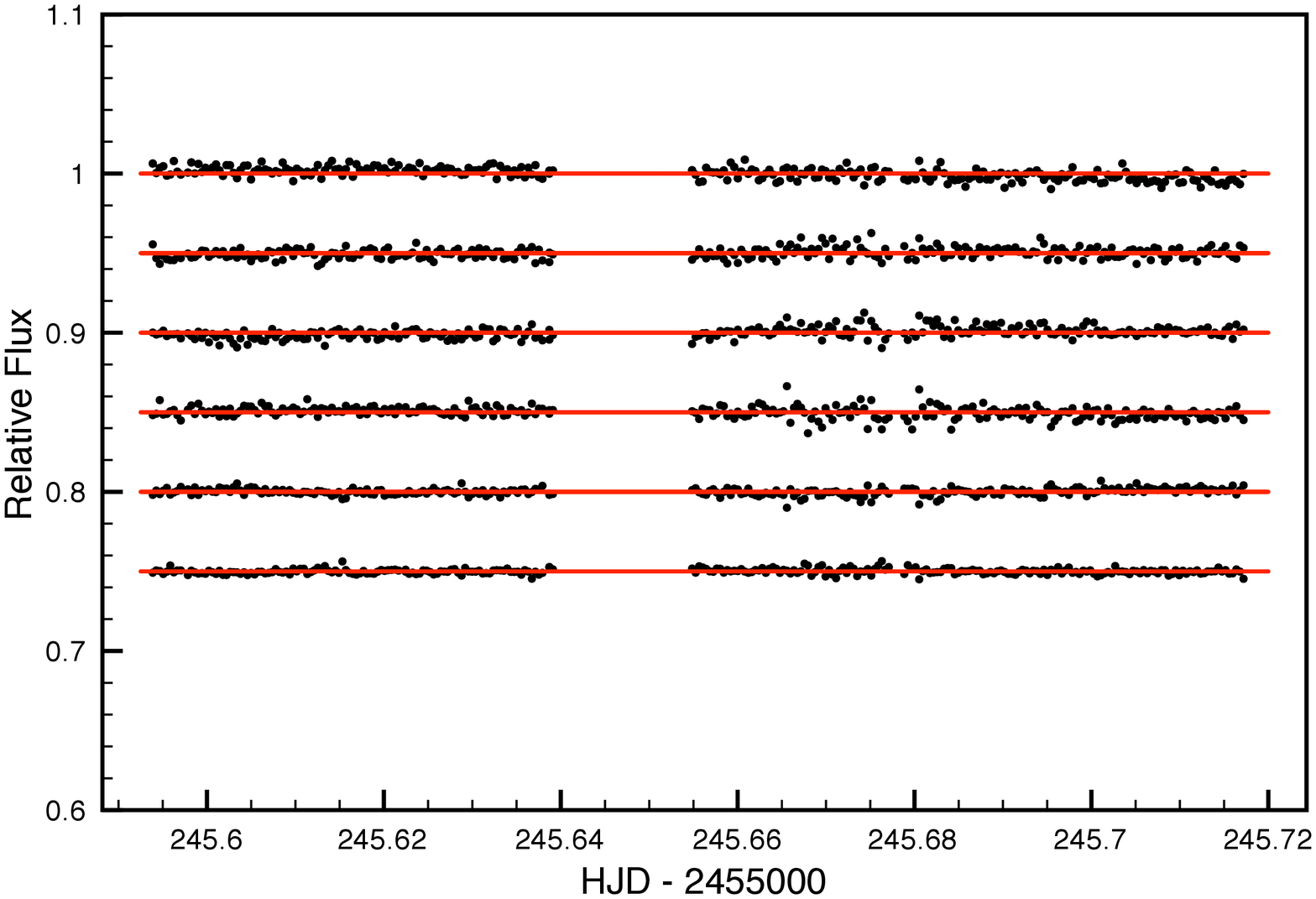}
\includegraphics[scale=0.35]{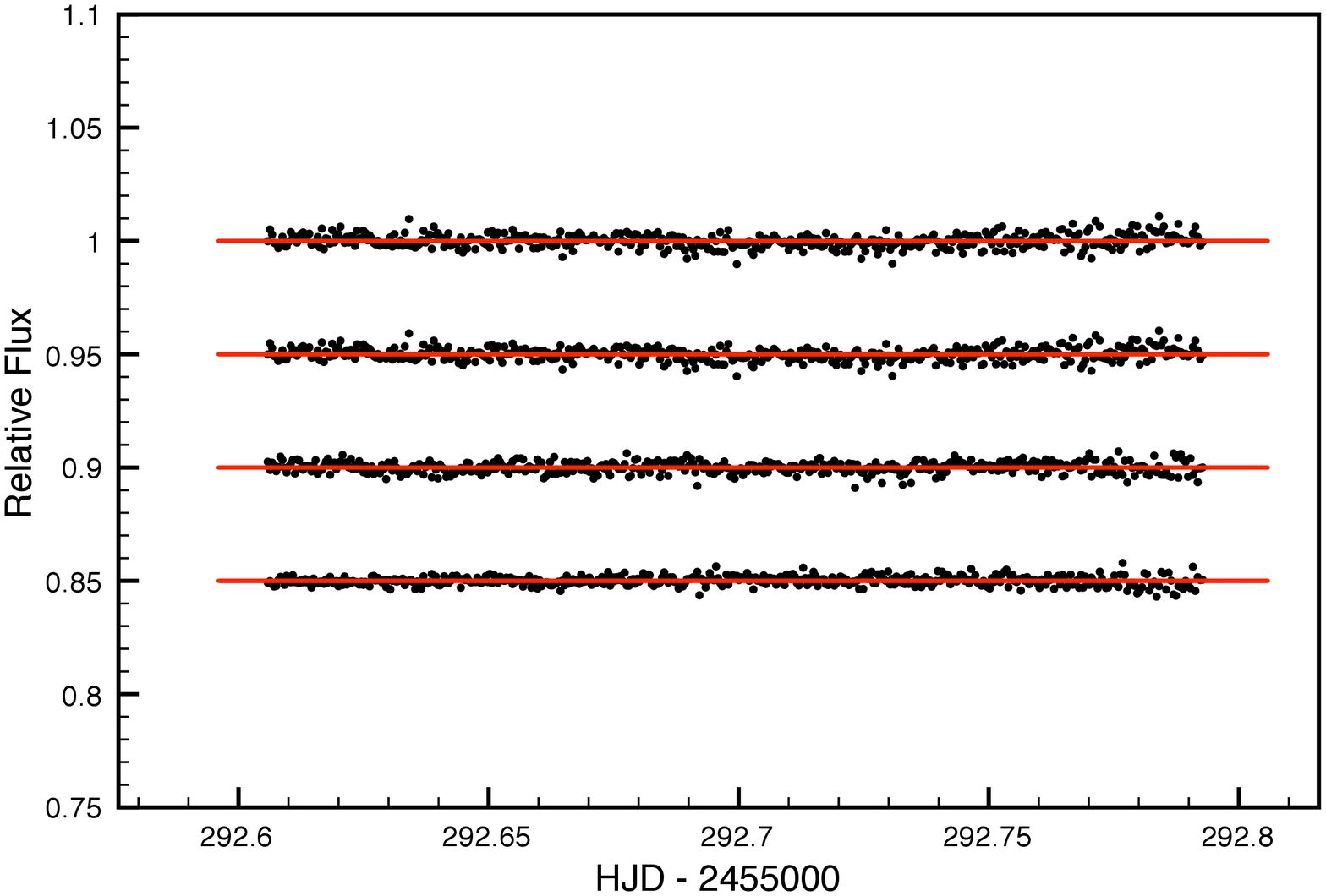}

\caption{Plot of the reference stars used to normalize each transit as a function of time. RMS scatter of these stars is approximately $\sim 2.0$ mmag, for February 14 (top) and April 05 (bottom). RMS scatter for February 18 (middle) was higher, at $\sim 2.5$ mmag, and partially obscured by clouds in the middle of the data set. The flux of each reference star was normalized by the average fluxes of the other three reference stars. The same four reference stars were used for all three nights, although two additional stars were used on the night of February 18th, as our RMS noise was reduced when we included them. This was not the case for these additional stars for the other two nights.
}
\label{RefStars}
\end{figure}

Each of the data images were bias-subtracted, flat-fielded, and bad pixel-cleaned in the usual manner. Aperture photometry was performed using an IDL pipeline utilizing the ``find" and ``aper" tasks available in the IDL Astronomy User's Library.\footnote{The IDL Astronomy User's Library is available at http://idlastro.gsfc.nasa.gov/} A 4$\arcsec$.3 aperture radius (corresponding to 10.0 pixels) was adopted because it produced the lowest scatter in the resultant lightcurves. Several combinations of reference stars were considered, but four were selected for the final reduction of our 14 February and our 05 April data because of the low RMS ($\sim 1.9-2.2$ mmag) scatter. These reference stars were all distributed a few arc minutes east of the target, but scattered both slightly North and slightly South of HAT-P-9. Two additional stars were used for our 18 February data because they reduced our residual RMS. We were unable to find sufficiently bright reference stars to the west of HAT-P-9 without increasing our RMS scatter. We show the time series light curve for each reference star (normalized to the average flux of the other reference stars) for each night in Figure \ref{RefStars}.

\begin{figure}[htp]
\centering
\includegraphics[scale=0.35]{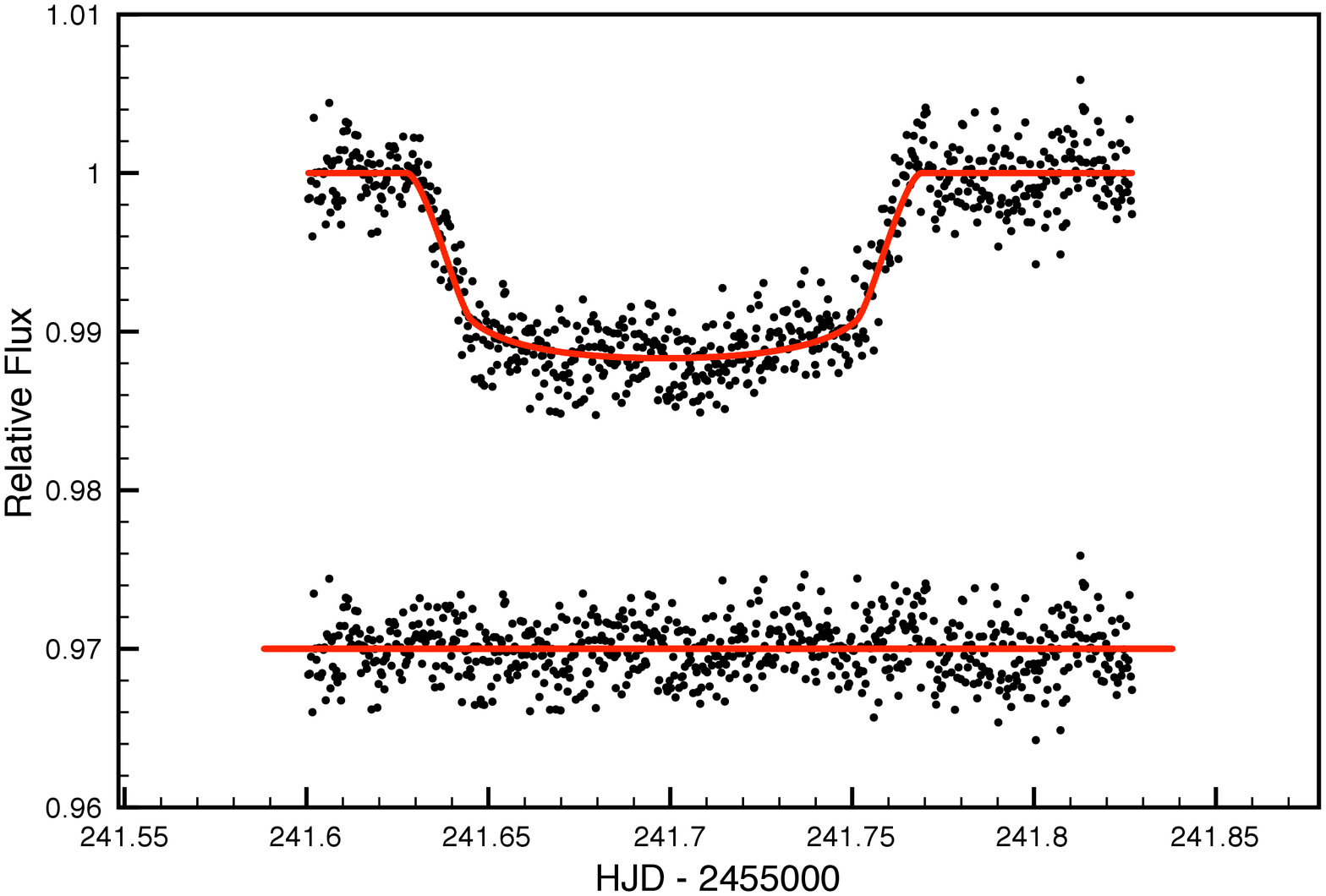}
\includegraphics[scale=0.35]{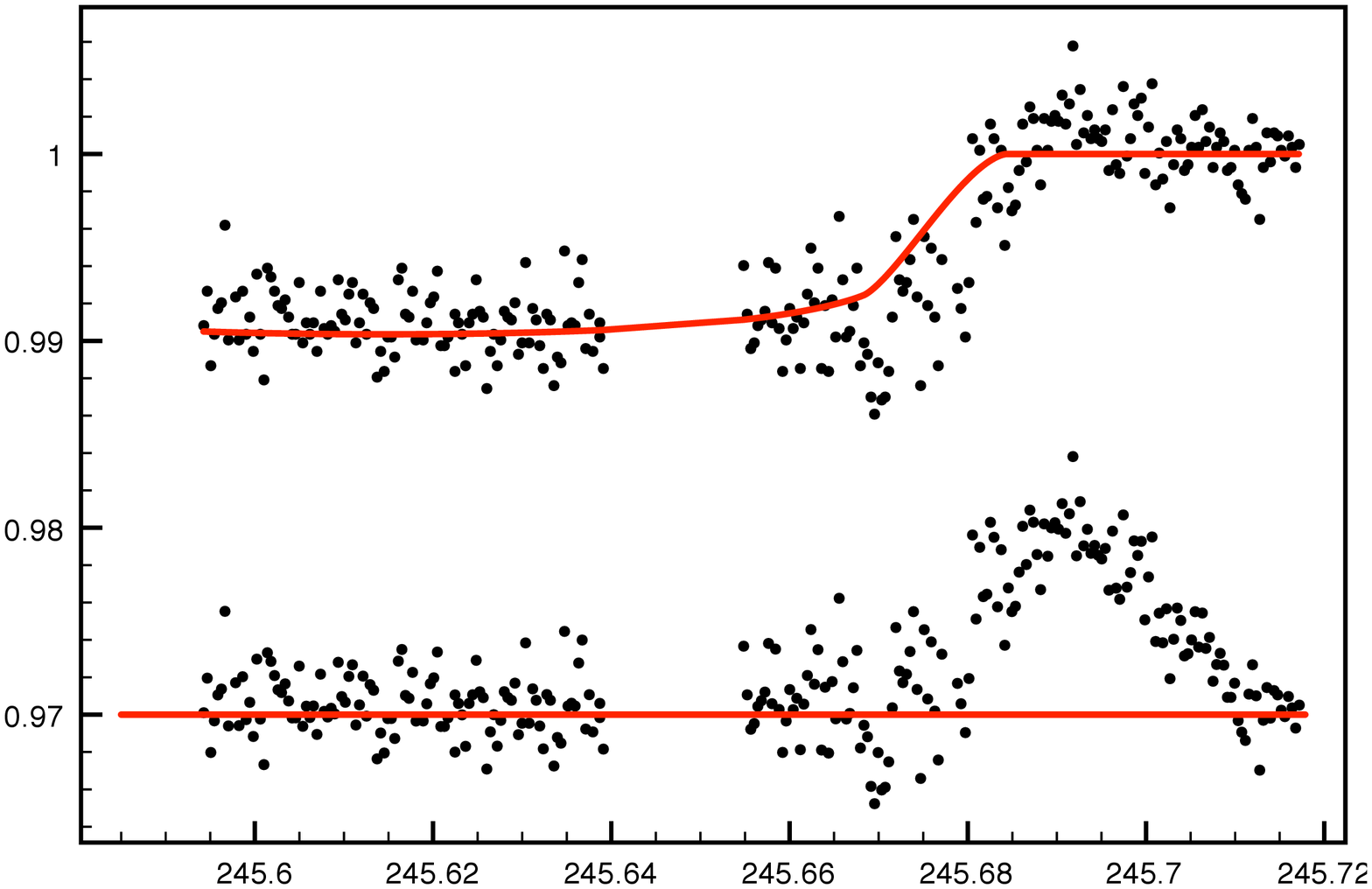}
\includegraphics[scale=0.35]{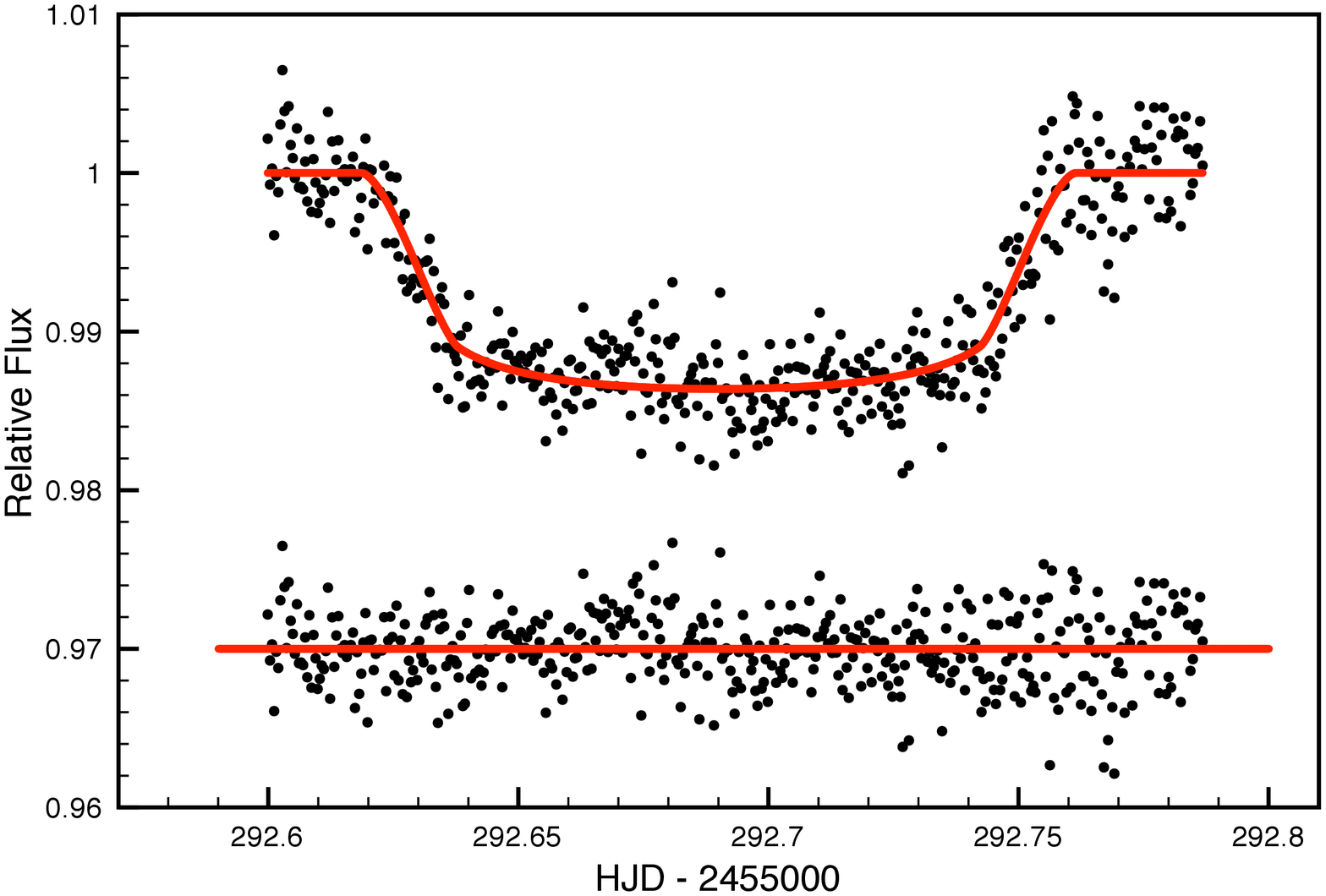}
\caption{Light curves of transits of HAT-P-9b taken on 14 February 2010 (top), 18 February 2010 (middle), and 05 April 2010 UT (bottom) at the University of Arizona's 1.55 meter Kuiper telescope on Mt Bigelow with the Mont4k CCD and the I filter. Standard deviation of the residuals (shown below the transit light curve) is 1.9 mmag forthe 14 February data, 2.5 mmag for the 18 February data, and 2.2 mmag for the 05 April data.  The transit was fit with the method of Mandel and Agol (2002), varying the central time of transit and planet to star radius ratio. Our transit time is in significant disagreement with the ephemeris of Shporer et al. (2008), and is $23.44 \pm 0.46$ minutes earlier than expected on 14 February, $23.74 \pm 1.44$ minutes earlier on 18 February, and $23.89 \pm 0.58$ min early on 05 April.}
\label{transit}
\end{figure}

We applied no sigma clipping rejection to the reference stars or HAT-P-9b; all datapoints were used in the analysis. The final light curve for HAT-P-9 was normalized by division of the weighted average of the reference stars. A slight (0.3\% maximum each night) parabolic correction was fit to the out of transit points and then used to normalize the entire light curve in order to correct for nonlinearities in the out of transit points due to differential color extinction between the bluer target and the redder reference stars. This correction was done only for our complete transit data sets, and not for the night of February 18, as we this correction requires out of transit data both before and after the transit to be effective. These corrections were small due to the low airmass of the observations, however slight errors in this correction may lead to a systematic error when determining the planet to star radius ratio. The unbinned residual time series in Figure \ref{transit} have a photometric RMS between 1.9 and 2.5 mmag, which is typical for the Mont4k on the 61-inch (Kuiper) telescope for our high S/N transit photometry pipeline (Dittmann et al. 2009a and b, Scuderi et al. (2010)). The higher 2.5 mmag RMS occured on the partially cloudy night of February 18.  

\section{Analysis}
We fit the transit light curve for February 13 and April 05 with the method prescribed by Mandel and Agol (2002), fitting for the central time of transit and planet to star radius ratio. We fit only for the central transit time of our partial February 17 data by taking our best fit February 13 model, and minimizing $\chi^2$ with respect to the central transit time. The fit and residuals from each fit are shown in Figure \ref{transit}. We note that we have not been able to fit the egress of our February 18 data, and in fact, the partial cloud cover has partially distorted the shape of egress. The shape has further been distorted because we were unable to correct for differential color extinction We have tried to account for this systematic effect with larger error bars in that night's central transit time. Linear and quadratic limb darkening coefficients in the I band were taken from Claret (2000) as 0.1874 and 0.3611 respectively. In order to estimate the errors for our fits, we generated and fit 1000 fake data sets by taking our measured data points and adding white noise with a standard deviation equal to the instrumental standard deviation of our data points for that night. The results of our fit, and those of Shporer et al. (2008), are shown in Table \ref{results}. In order to estimate the effect of our parabolic differential extinction correction on the estimate of transit depth ($R_p/R_*$)$^2$, we estimated the systematic error as  $30 \%$ the maximum parabolic correction, and folded that into our error from the Monte Carlo simulations. 

We find a central transit time for HAT-P-9b of $2455241.69832 \pm 0.00032$ HJD for the 14 February 2010 transit. Projecting outward 210 orbits with the ephemeris of Shporer et al. (2008), we find an expected time of $2455241.71460$ HJD. This means that our transit time is $23.44 \pm 0.46$ minutes earlier than expected from this ephemeris. Furthermore, we find central transit times of HJD $2455245.6312 \pm 0.0010$ and HJD $2455292.69006 \pm 0.00036$ for our February 18 and April 05 transit data sets, which are $23.74 \pm 1.44$ min and $23.89 \pm 0.58$ min early, respectively. All three of our transits are entirely inconsistent with the ephemeris of Shporer et al. (2008).

\section{Discussion}
The observations of Shporer et al. (2008) include 4 transits over 15 orbits between 13 November 2007 and 11 January 2008. The proximity in time between these observations naturally limit the ability to obtain a precise ephemeris for our observations 210 orbits into the future. Taking addtional observations many orbits later can drastically improve the fit and allow for better prediction of future transits. Indeed, in order to eliminate our large, $ \approx -24$ minute deviation from the expected transit time, we must only shift the period of Shporer et al. (2008) $6.5$ seconds ($2\sigma$) downwards to P = $3.922814 \pm 0.000002$ days. Therefore, we predict for future transits:

\begin{equation}
T_c = 2454417.9077 + 3.922814 * E \textrm{    HJD}
\end{equation}

\begin{figure}[htp]
\centering
\includegraphics[scale=0.35]{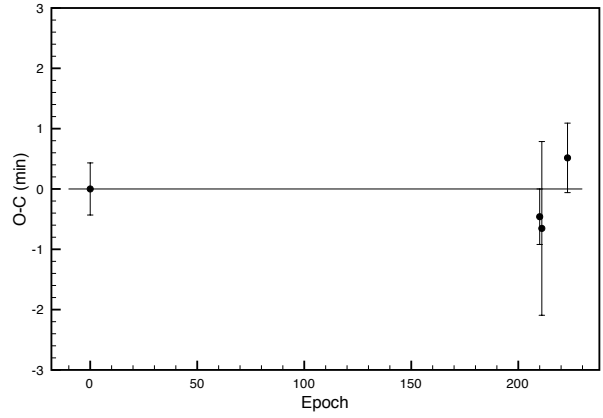}
\caption{Observed minus calculated (OC) plot for all currently available transit data for the HAT-P-9 system using our $3.922814 \pm 0.000002$ day period. All currently available data is consistent with a linear period, and this lack of large TTVs doesn't suggest any additional gravitational perturbers in the HAT-P-9 system.}
\label{OC}
\end{figure}

Where E represents the integer epoch counting from Shporer et al. (2008)'s measurement. We note that we have improved the measurement of the period of this system by one order of magnitude. Furthermore, with this orbital period, it now becomes experimentally viable to probe the HAT-P-9 system for future transit timing variations (TTVs). We have constructed an observed minus calculated plot (see Figure \ref{OC}), to investigate any possible nonlinearities in the currently observed transit times. To date, we find no evidence for any nonlinearities in the transit times of HAT-P-9b greater than $\sim 1$ min, suggesting that if other bodies were to exist in this system, they would not be in close orbital resonances with HAT-P-9b. However, we note that if it turns out that this ephemeris cannot predict future transits in 2010 and beyond, then that will be substantial evidence for nonlinear TTVs, and a possible indicator of additional bodies in the system.

While we have obtained two different measurements for the planet to star radius ratio for HAT-P-9b, due to the blue nature of the star and the redness of the reference stars, we found it necessary to apply a parabolic correction to correct for differential color extinction as the airmass of the observations changed. This introduced a possible systematic shift in the transit depth depending on the accuracy of this correction. We have estimated this systematic error as approximately 30\% of the maximum parabolic correction. When this is taken into account, our observed $R_p/R_*$ ratios are within a few $\sigma$ of the value from Shporer et al. (2008), and we find no significant difference between these values. 

\section{Conclusions}
Here we have investigated three follow-up transits of HAT-P-9b in order to obtain a more accurate ephemeris. We have found that our transits occurred $\sim 24$ minutes earlier than expected from the ephemeris of Shporer et al. (2008). However, due to the time between our observations and those of Shporer et al. (2008), we find that only a $2\sigma$ decrease in the period is sufficient to explain all currently available observations. We have found a new period of $3.922814 \pm 0.000002$ days. Future observations will be able to determine whether our new period is sufficient to describe the system, and any future deviations from this ephemeris may indicate the presence of additional bodies in the system. 

\acknowledgments

JD and LMC are supported by a NSF Career award and the NASA Origins program.

{\it Facilities:} \facility{Kuiper 1.55m}. \\
\\ {\bf References}\\\
\bibliography
BBakos, G. \'A., Torres, G. P\'al, A., et al., 2010, ApJ, 710, 1724 \\
Ballard, S. Christiansen, J.L, Charbonneau, D., et al. 2010, AAS, 425 \\
Baraffe, I., Chabrier, G., and Barman, T. 2008, A\&A, 482, 315 \\
Burrows, A., Hubeny, I., Budaj, J., \& Hubbard, W. B. 2007, ApJ, 661, 502 \\
Butler, R.P., Vogt, S.S., Marcy, G.W., et al. 2004, ApJ, 617, 580 \\
Claret, A., 2000, A\&A, 363, 1081 \\
Deming, D. Harrington, J., Laughlin, G., et al., 2007, ApJ, 667, L199 \\
Dittmann, J.A., Close, L.M., Green, E.M., Scuderi, L.J., Males, J.R. 2009a, ApJ, 699, L48.\\
Dittmann, J. A., Close, L.M., Green, E.M., Fenwick, M. 2009b, ApJ, 701, 756 \\
Fortney, J.J., Marley, M.S., and Barnes, J.W., 2007, ApJ, 659, 1661 \\
Hansen, B. M. S. and Barman, T. 2007. ApJ, 671, 861 \\
Haghighipour, N., Steffen, J., \& Agol, e. 2008, ApJ, 686, 621 \\
Mandel, K., \& Agol, E. 2002, ApJ, 580, L171\\
Maness, H.L, Marcy, G.W., Ford, E.B., et al., 2007, Pub. Ast. Soc. Pac., 119, 90 \\
Pont, F., Gilliland, R.L, Knutson, H., et al. 2009, MNRAS, 393, L6 \\
Scuderi, L.J., Dittmann, J.A., Close, L.M, et al. 2009, ApJ, in press (arXiv:0907.1686) \\
Shporer, A., Bakos, G., Bouchy, F., Pont, F. et al. 2008, ApJ, 690, 1393. \\
\clearpage

\begin{table}
\begin{center}
\caption{Parameters of the HAT-P-9 system}\label{results}

\begin{tabular}{crrr}
\tableline\tableline
  Parameter [units] & Value & Reference \\
  \tableline
 $R_p / R_*$ & $0.1083 \pm 0.0005$ & Shporer et al. (2008) \\
 $R_p / R_*$ & $0.1031 \pm 0.0013$ & This Work$^a$ \\
 $R_p / R_*$ & $0.1109 \pm 0.0014$ & This Work$^b$ \\
 $T_c$ (HJD-2,400,000) & $54417.9077 \pm 0.0003$ & Shporer et al. (2008) \\
 $T_c$ (HJD-2,400,000) & $55241.69832 \pm 0.00032$ & This work \\
 $T_c$ (HJD-2,400,000) & $55245.621 \pm 0.0010$ & This work \\
 $T_c$ (HJD-2,400,000) & $55292.69558 \pm 0.00036$ & This work \\
 P (days) & $3.92289 \pm 0.00004$ & Shporer et al. (2008) \\
 P (days) & $3.922814 \pm 0.000002$ & This work \\
\tableline
\label{results}
\end{tabular}

\tablenotetext{a}{Best fit value from the February 14 observing night}
\tablenotetext{b}{Best fit value from the April 05 observing night}

\end{center}
\end{table}

\clearpage

\clearpage

\clearpage

\clearpage

\end{document}